\documentclass[a4paper,11pt]{article}
\usepackage{pos}

\usepackage{lineno}
\usepackage{array}
\usepackage{makecell}
\usepackage{caption}
\usepackage{subcaption}
\usepackage{multirow}
\usepackage{lineno}
\usepackage{comment}
\usepackage{tabularx}
\usepackage[ragged]{sidecap}
\usepackage{siunitx,upgreek}
\usepackage{lineno}
\usepackage{xcolor}

\title{Performance of the AstroPix Prototype Module for the Barrel Imaging Calorimeter at the ePIC Detector and in Space-Based Payloads}
\author*[a]{Bobae Kim}
\author[b]{Regina Caputo}
\author[a]{Manoj Jadhav}
\author[a]{Sylvester Joosten}
\author[b]{Adrien Laviron}
\author[c]{Richard Leys}
\author[a]{Jessica Metcalfe}
\author[c]{Nicolas Striebig}
\author[b]{Daniel Violette}
\author[a]{Maria Żurek}

\affiliation[a]{Argonne National Laboratory \\
9700 S Cass Ave, Lemont, 60439, IL, USA}
\affiliation[b]{NASA Goddard Space Flight Center \\
8800 Greenbelt Rd, Greenbelt, 20771, MD, USA}
\affiliation[c]{Karlsruhe Institute of Technology \\
Hermann-von-Helmholtz-Platz 1, Karlsruhe, D-76344, Baden-Württemberg, Germany}
\emailAdd{bobae.kim@anl.gov}
\abstract{
AstroPix is a high-voltage CMOS (HV-CMOS) monolithic active pixel sensor originally developed to enable precision gamma-ray imaging and spectroscopy in the medium-energy regime ($\sim$100$~\unit{keV}$--100$~\unit{MeV}$) based on the groundwork laid by ATLASpix and MuPix. It features a 500$~\unit{\micro\meter}$ pixel pitch, in-pixel amplification and digitization, and low power consumption ($\sim$3-4$~\unit{mW/cm}^2$), making it scalable for large-area, multilayer telescope detector planes. The detectors have a designed dynamic range of 25$~\unit{keV}$ to 700$~\unit{keV}$.

With these features, AstroPix meets the requirements of future space-based high-energy telescopes and the imaging layers of the Barrel Imaging Calorimeter (BIC) in the Electron-Proton/Ion Collider (ePIC) detector at the future Electron-Ion Collider (EIC). For the space-based payload, AstroPix is being integrated into sounding rocket and balloon payloads to demonstrate the technical readiness of the devices. For BIC, AstroPix-based imaging layers interleaved within the lead/scintillating-fiber (Pb/SciFi) sampling calorimeter provide granular shower imaging, enabling key performance features such as electron/pion or gamma/neutral pion separation.

As part of the ongoing detector R\&D efforts, we have been testing various AstroPix\_v3 configurations: the single chip, a quad-chip assembly, a three-layer stack of quad chips, and a 9-chip module that represents the smallest prototype unit of the BIC imaging layer. This presentation will highlight recent performance test results from these AstroPix detector configurations.
}

\FullConference{33rd International Workshop on Vertex Detectors (VERTEX2025)\\
25-29 August 2025\\
University of Tennessee, Knoxville, USA\\}

\begin{document}
\maketitle
\section{Introduction}
\label{sec:intro}

\begin{figure}[h!]
    \centering
    \includegraphics[width=0.285\linewidth]{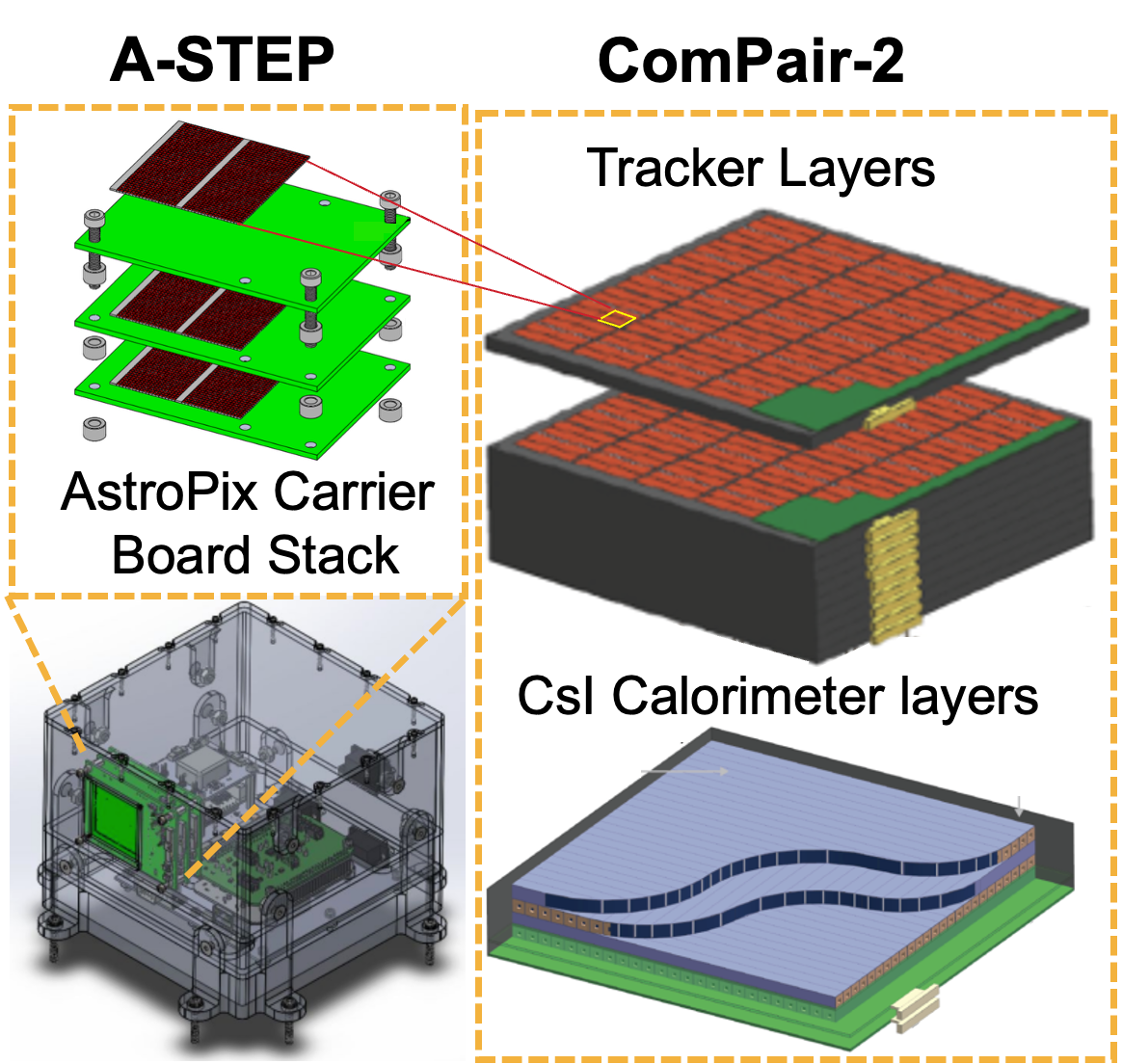}
    \includegraphics[width=0.55\linewidth]{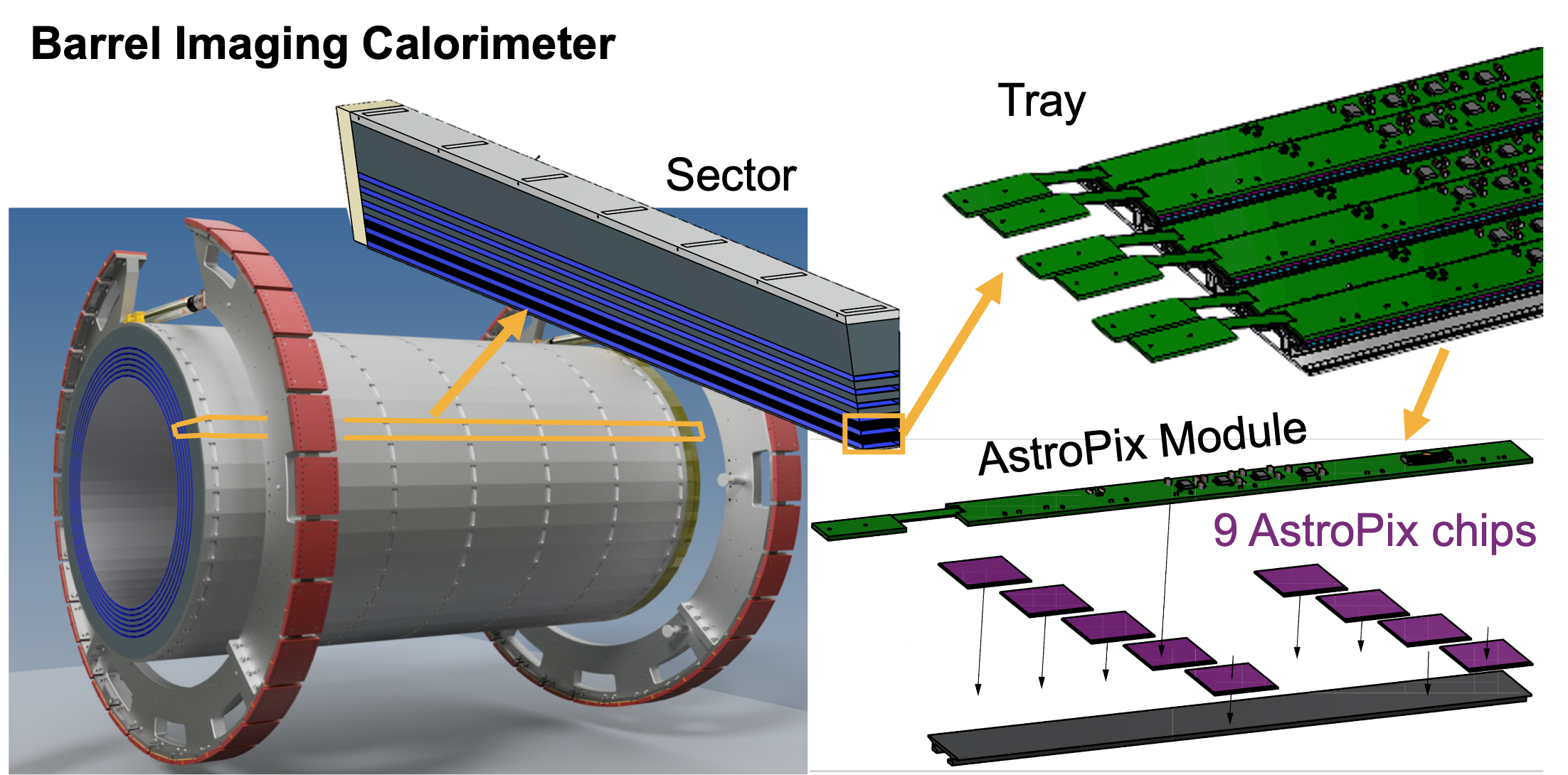}

    \caption{Detector structure of A-STEP and ComPair-2 missions at NASA Goddard Space Flight Center (left) and Barrel Imaging Calorimeter (right).}
    \label{fig:detstr}
\end{figure}

%A novel High Voltage-CMOS (HV-CMOS) monolithic active pixel sensor named AstroPix has been developed for the proposed All-sky Medium-Energy Gamma-ray Observatory eXplorer (AMEGO-X) mission concept.
%A medium-energy (100 keV to 100 MeV) gamma-ray telescope will deepen our understanding of various astrophysical phenomena and processes by identifying and characterizing gamma rays from extreme explosions and astrophysical particle accelerators~\cite{caputo2022amegox}.
A novel High-Voltage CMOS (HV-CMOS) monolithic active pixel sensor named AstroPix has been developed for the proposed All-sky Medium-Energy Gamma-ray Observatory eXplorer (AMEGO-X) mission concept, a medium-energy (100 keV–100 MeV) gamma-ray telescope designed to deepen our understanding of astrophysical phenomena by identifying and characterizing gamma-rays from explosions and high-energy particle acceleration in the universe~\cite{caputo2022amegox}.
%such as compact object mergers, active galactic nuclei, and gamma-ray bursts, by reconstructing both Compton-scattering and pair-production photon events in next-generation observatories.
The ComPair-2 balloon mission serves as a flight-like prototype for the AMEGO-X, designed to validate detector design and performance in a near-flight environment~\cite{Caputo:2024kdg}.
It features a 10-layer tracker with 95 quad-chips per layer and a 4-layer calorimeter composed of 24 CsI bars coupled to SiPMs on each side, as shown in Fig.~\ref{fig:detstr}.
To further validate the tracker subsystem layers of a Compton and pair (ComPair-2) gamma-ray telescope in a relevant space-like environment, the AstroPix Sounding Rocket Technology dEmonstration Payload (A-STEP) is being developed to demonstrate a three-layer stack of AstroPix\_v3 quad-chips on a sub-10 minute sub-orbital rocket flight~\cite{violette2024astep}.

%In parallel, AstroPix is also being developed for applications in nuclear physics experiments at the upcoming Electron-Ion Collider (EIC).
%EIC solves the fundamental questions in nuclear physics . BIC is essential component in the ePIC detector, tailored to the first EIC experiment.
%In terms of nuclear physics, the ePIC experiment is the first experiment at the Electron-Ion Collider (EIC), dedicated to exploring fundamental questions in nuclear physics, and designed to deliver the EIC physics program.
The Electron-Ion Collider (EIC) will explore matter’s inner structure to provide unprecedented 3D imaging of protons and nuclei and study gluon dynamics, including proton spin, saturation, and confinement.
%The EIC solves the fundamental questions in nuclear physics. 
The EIC will provide a unique beam environment where polarized electrons collide with polarized protons and light nuclei, as well as unpolarized heavier nuclei, at asymmetric beam energies, offering center-of-mass energies of 20–140~GeV and luminosities up to $10^{34}$~\unit{\cm^{-2}\per\second}.
The Barrel Imaging Calorimeter (BIC) is an essential component of the ePIC detector for the first EIC experiment.
It serves as the barrel electromagnetic calorimeter, combining lead-scintillating fiber sampling calorimeter layers (Pb/SciFi layers) with AstroPix-based imaging layers to provide precise energy and shower profile measurements.
AstroPix layers embedded in the front half of the calorimeter will provide fine-grained three-dimensional shower imaging, enabling key performance capabilities such as $e^-$/$\pi^\pm$ and $\gamma$/$\pi^0$ separation, which are critical for Deep Inelastic Scattering (DIS) measurements~\cite{abdulkhale2022eicyellow, Klest2025}.

As shown in Fig.~\ref{fig:detstr}, BIC consists of 48 sectors, each containing an AstroPix imaging layer and Pb/SciFi layers. The AstroPix layer is composed of trays, and each tray contains 72--84 modules.
Therefore, the AstroPix module with a nine daisy-chained chips constitutes the basic unit of the BIC imaging layer, while the quad-chip serves as the fundamental unit for the A-STEP and ComPair-2 payloads.
The present study focuses on validating the performance of these prototype modules and demonstrating their operational feasibility.
%As shown in Fig.~\ref{fig:detstr}, the detector structure of the BIC is shown together with A-STEP and ComPair-2, both of which are NASA space-based missions.
%-----------
% AstroPix\_v3 
%-----------
\section{AstroPix\_v3}
Since 2019, AstroPix\_v3 has been developed and tested from version 1 to version 4~\cite{steinhebel2022astropix,SUDA2024169762,steinhebel2025astropix,striebig2024astropix4,suda2026astropix4}. %이전버전 결과들.
AstroPix\_v3 is the first full-size chip and has been selected for the A-STEP, ComPair-2, and BIC projects, serving as the current prototype.
It has the reticle $\sim$2$\times$2 \unit{\cm^2} size and features a 35 $\times$ 35 pixel matrix with a pixel pitch of 500~\unit{\micro\meter}. 
AstroPix\_v3 was fabricated using TSI Semiconductors’ 180~nm process with a substrate resistivity of 200--400~$\Omega \cdot$~\unit{\cm} and thickness of 725~\unit{\micro\meter}.
The design includes row and column hit buffers and supports an OR’ed rows-and-columns readout scheme. 
It has in-pixel amplification and low power consumption ($\sim$3--4$~\unit{mW/cm}^2$), making it scalable for large-area, multilayer tracker telescope detector planes. 
The chip operates with a 2.5~MHz clock for a time of arrival (ToA) with an 8 bit counter and up to 200~MHz clock for Time-over-Threshold (ToT) measurement with a 12 bit counter.
Each hit in the row/column readout generates a data packet containing the payload size, layer, chip ID, row/column number, ToA, and ToT.
AstroPix\_v3 employs streaming readout and self-triggering at the pixel level, where each pixel autonomously records hits above threshold.
%, MIP response and depletion depth measurement using 120 GeV proton~\cite{SUDA2024169762, steinhebel2025astropix}.%MyBeamTestPaper!
%Previous studies of energy resolution and dynamic range of single-chip have already been conducted~\cite{SUDA2024169762}, with Cd-109, Ba-133, Am-241, and Co-57 sources covering the energy range from 22.2 keV to 122 keV.
%v3 basic (general?) performance : Amanda's paper
%beam test result MIP response, depletion depth : my paper
%The dynamic range of AstroPix\_v3 single-chip was confirmed from 25 to 200 keV, although the final target of AstroPix5 for the production is 25--700 \unit{keV}, as required for the NASA space-based missions. 
%44\% of pixels meet the energy resolution requirement of 10\% at 59.5 keV with a median full-width half-maximum of 6.2 keV (10.4\%).
%In addition, 92.4\% of pixels achieved the low-energy floor requirement of 25 keV sensitivity, also required for the BIC. 

\begin{figure}[h!]
    \centering
    \includegraphics[width=1\linewidth]{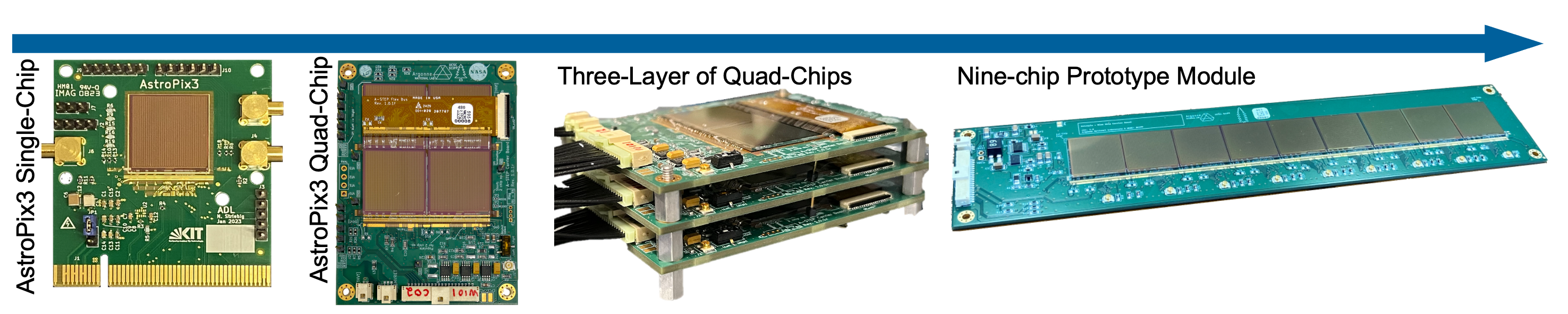}
    \caption{AstroPix\_v3 configurations: stepwise progression from single-chip to nine-chip prototype module.}
    \label{fig:conf}
\end{figure}
Starting from the single chip, each configuration has been tested with progressively increasing scale, as shown in Fig.~\ref{fig:conf}.
The single chip has $\sim$2~\unit{cm} $\times$ 2~\unit{cm}; the quad-chip is a 2 $\times$ 2 array of single chips (sensor size $\sim$4~\unit{cm} $\times$ 4~\unit{cm}); and the nine-chip module is a 1 $\times$ 9 linear array of single chips.
For the single-chip module, bond pads located along the lower edge of the chip digital periphery were directly wire-bonded to a rigid carrier PCB. 
For the quad-chip module, the bond pads along the lower edge of the digital periphery of the top chips were wire-bonded to a flexible PCB bus bar, which was mounted on the top chips to provide power and communication.
The quad-chip and nine-chip modules were subsequently connected in a daisy-chain configuration and read out via a SPI interface.
%The nine-chip prototype module used a similar PCB based on single chip or quad-chip carrier board and it is a mock-up of AstroLinx, which represents the final rigid-flex PCB for the module electronics.

%Daisy Chain readout - pass hits to next chip through QSPI ?
% SPI I/O daisy chained - chip-to-chip signal transfer 
% - signals are digitized & routed out to the neighboring chip using 5 SPI lines via wire bond
% - Power/Logic I/O distribution on the module (through a bus tape)
%------------
%Test Results
%------------
\section{Test Results}
\begin{figure}[h!]
    \centering
    \includegraphics[width=0.55\linewidth]{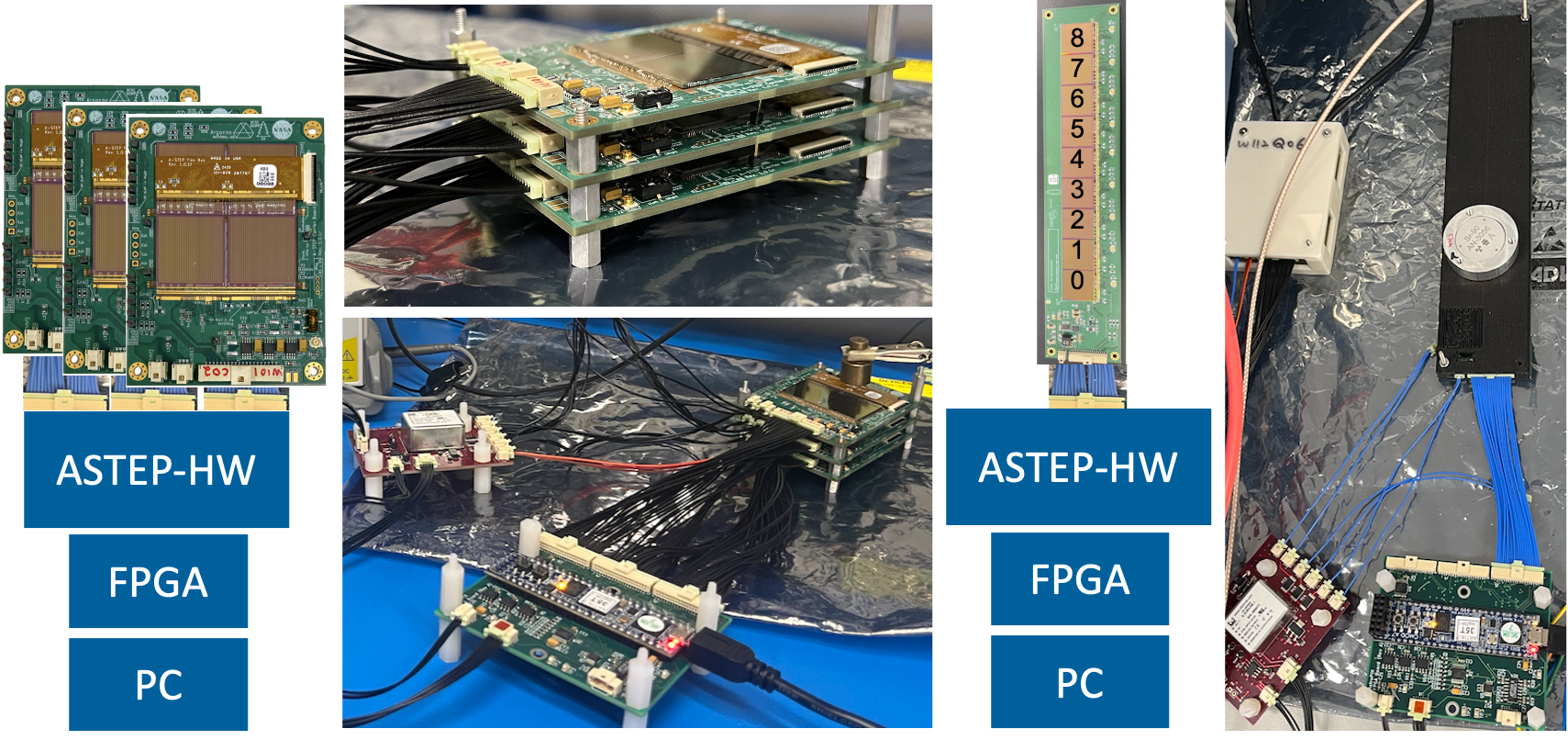}
    \caption{Pictures of the bench test setup with AstroPix\_v3 configurations.}
    \label{fig:bench}
\end{figure}

The main goal of the tests with various AstroPix\_v3 configurations is to validate their operation—including basic device functionality such as DAQ communication and daisy-chain readout—as well as performance aspects such as noise level, chip uniformity, pixel-to-pixel variation, and rate capability, evaluated through injection tests, cosmic-ray tests, and source tests.
%For the multi-chip setup, uncalibrated chips were used, and the tests primarily aimed to validate the operation with each configuration as well as the software and firmware currently under development.
Based on the performance result of single-chip tests, a bias voltage of -150 V and a global threshold of 200 mV were applied to all AstroPix\_v3 configurations with the current setups.
The performance of the AstroPix\_v3 single-chip, including noise characteristics, injection response, energy resolution, calibration, and dynamic range, has been reported in previous studies~\cite{SUDA2024169762, steinhebel2025astropix}. 

For data acquisition, the A-STEP hardware (A-STEP HW) developed as a sounding rocket payload, together with a commercial Cmod A7 Artix-7 FPGA Module, was used for the current performance evaluation, as depicted in Fig.~\ref{fig:bench}. 
This system includes an HV bias board that supplies up to –150 V to each AstroPix\_v3 carrier board and supports configurations of up to three layers of quad-chips or nine-chip modules.
Testing was performed with a quad-chip, a three-layer stack of quad-chips, and a nine-chip module.
In this section, representative test results with AstroPix\_v3 multi-chip configurations are highlighted.

\subsection{Quad-chip}
%w123q10 result.
%ssh daq@192.168.0.221
%->didn't work
\begin{figure}[h!]
    \centering
    \includegraphics[width=0.295\linewidth]{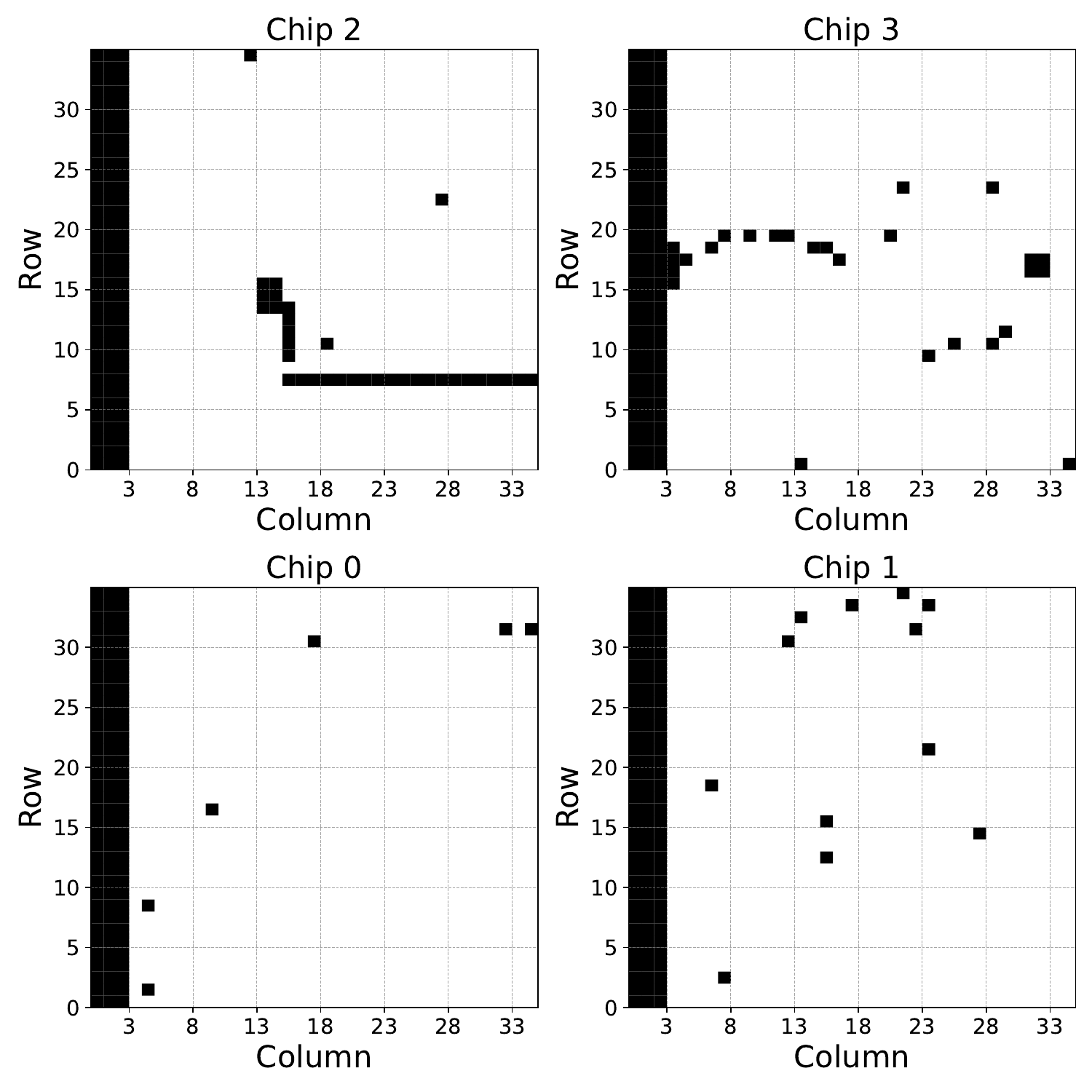} %maksmap_w123q10.pdf
    \includegraphics[width=0.34\linewidth]{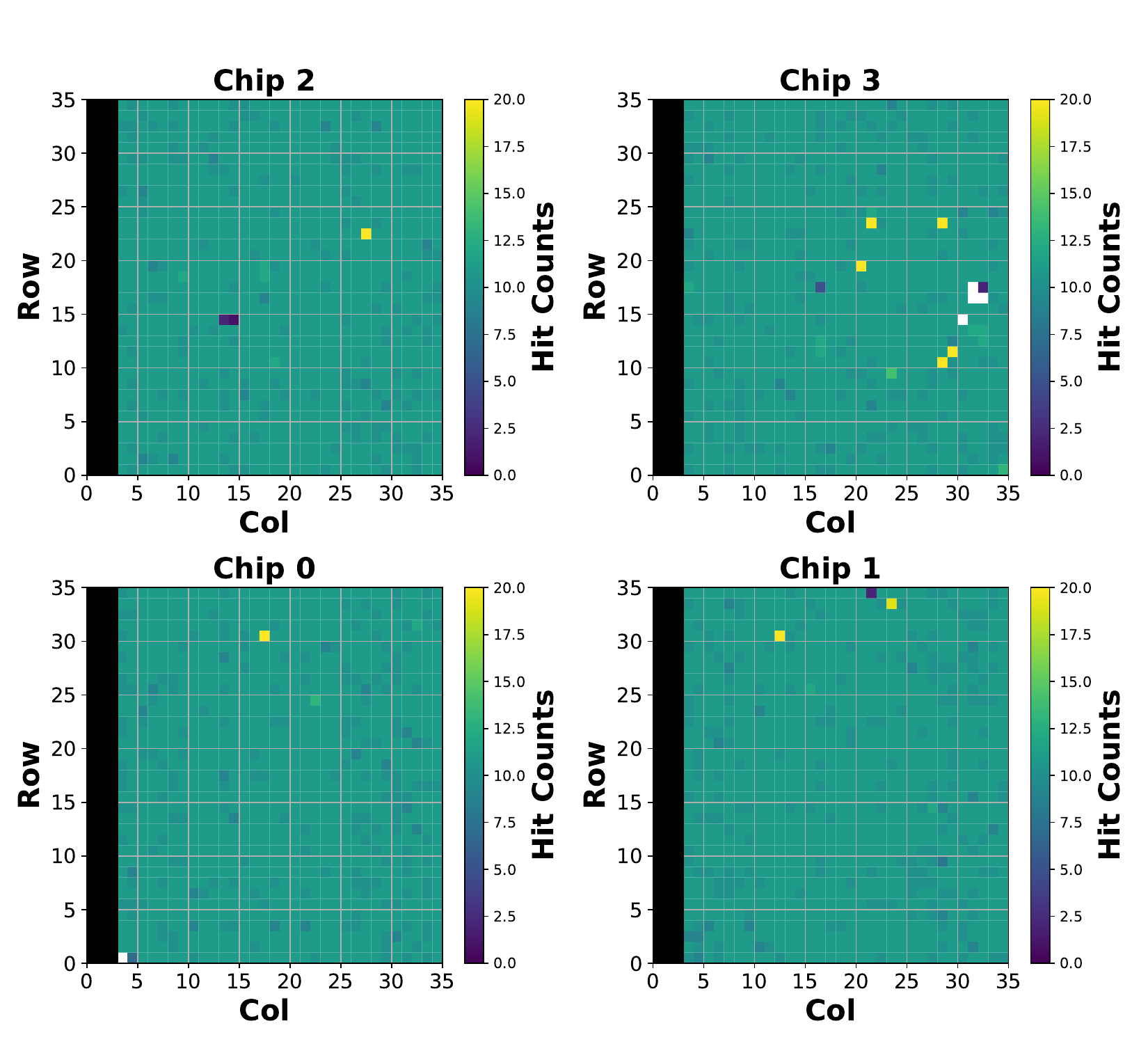}
    \includegraphics[width=0.34\linewidth]{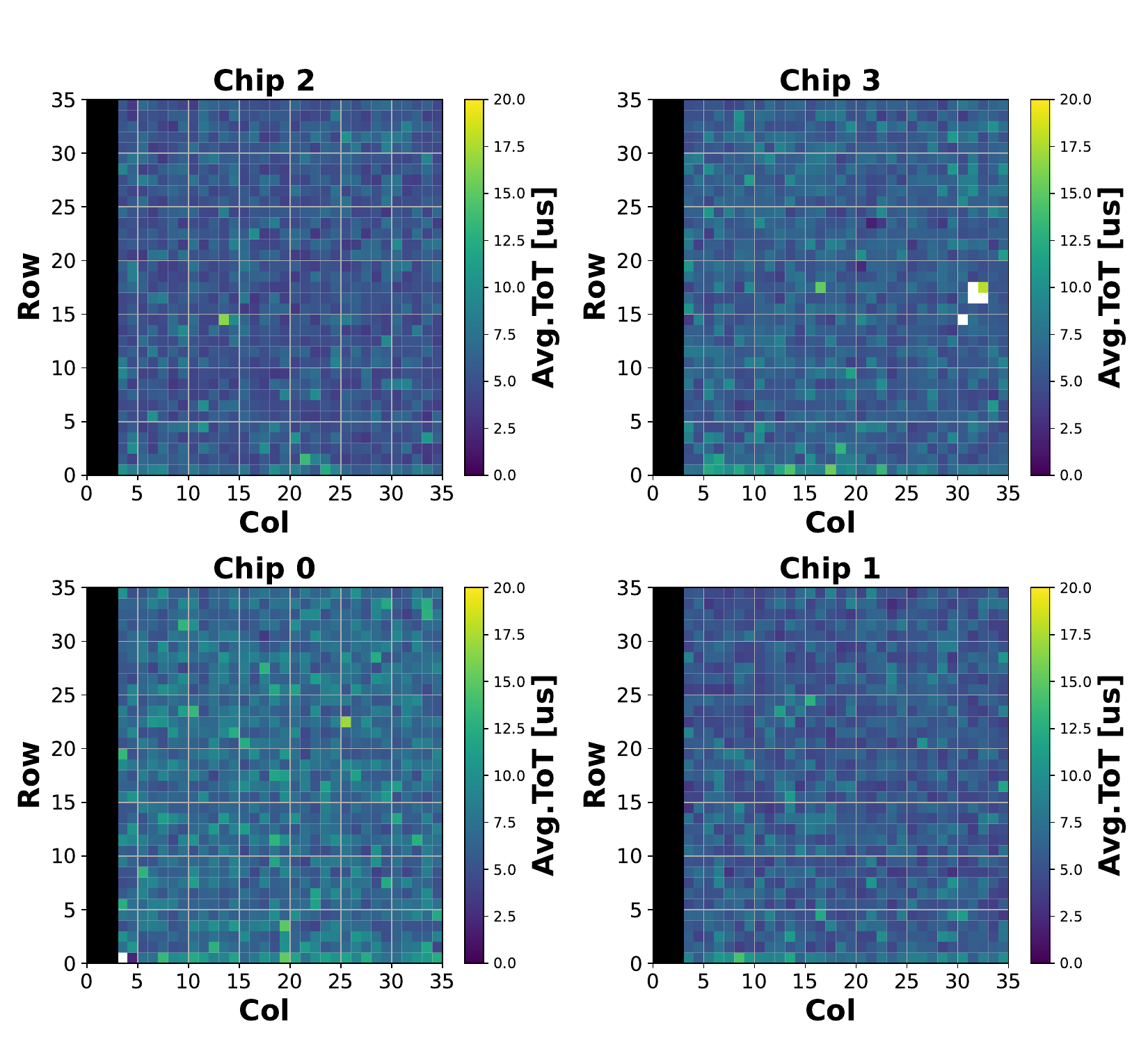}
    \caption{Mask map of the quad-chip module (left) and results of the injection test: hit map (middle) and ToT map (right). Each plot corresponds to one chip in the 2$\times$2 array.}
    \label{fig:qnoise}
\end{figure}
As shown in Fig.~\ref{fig:qnoise}, the mask maps of a quad-chip were obtained by measuring noise-triggered counts per chip, and pixels with a noise rate greater than 2~Hz were masked and shown in black.
The first three columns were additionally masked because they use a different comparator design, which causes larger noise due to higher amplification.
An active pixel yield of 99\% was achieved in the two bottom chips, while a slightly lower yield of $\geq$97\% was observed in the two top chips due to the noise induced by a digital communication line on the flexible PCB bus bar, which appears as a straight line in chip~2 and as a curved line near the middle-right side of chip~3.
This effect is present only under multi-chip activation, appearing when three or four chips are enabled simultaneously, and is not observed in per-pixel readout tests.
%noise induced by a digital communication line (straight lines in chip 2 and curved line in chip 3

%\begin{figure}[h!]
%    \centering
%    \includegraphics[width=0.47\linewidth]{figs/q_inj_hitmap.png}
%    \includegraphics[width=0.47\linewidth]{figs/q_inj_totmap.png}
%    \caption{Injection test result of quad chip: Hit map (left) and ToT map (right)}
%    \label{fig:qsr90}
%\end{figure}
%The injection test was performed to verify pixel response uniformity and variation using the same injection voltage per pixel, evaluated through hit maps and ToT maps.
The injection test was performed to evaluate pixel-to-pixel response uniformity and variation, using the same injection voltage per pixel, based on the hit and ToT maps.
Hit maps, presenting the hit count per pixel, and ToT maps, presenting the average ToT value per pixel for each chip of the quad-chip, are shown in Fig.~\ref{fig:qnoise}.
Expected total counts per pixel is about 10--11 counts and hit maps shows good uniformity across the chips except for a few pixels.
Some pixels did not respond properly to the injection, and a few masked pixels show abnormal hit counts.
ToT maps present mean ToT values for each pixel and shows both chip uniformity and pixel-to-pixel response variation.
Most pixels exhibit mean ToT values generally in the range of 5--7~\unit{\micro s}, and their ToT distributions are well described by a Gaussian function with respect to the injection voltage.
These results are comparable with AstroPix single-chip pixel-to-pixel variation of 20--30\%.

%\begin{figure}[h!]
%    \centering
%    \includegraphics[width=0.75\linewidth]{figs/q_sr90_0.png}
%    \includegraphics[width=0.45\linewidth]{figs/q_sr90_1.png}
%    \includegraphics[width=0.45\linewidth]{figs/q_sr90_2.png}
%    \includegraphics[width=0.45\linewidth]{figs/q_sr90_3.png}
%    \caption{Source test result using Sr90 of quad chip: hit map (left) and ToT map (right)}
%    \label{fig:qsr90}
%\end{figure}
%The source test with Sr-90 was conducted to verify source response, evaluated via hit maps and ToT maps.
%The hit map aligns well with the source position.
%The ToT distribution is well-described by a Landau convoluted with a Gaussian function.
%comparable with ToT distribution of AstroPix single-chip result.

\subsection{Three-layer stack of quad-chips as the A-STEP payload}
\begin{figure}[h!]
    \centering
    \includegraphics[width=0.8\linewidth]{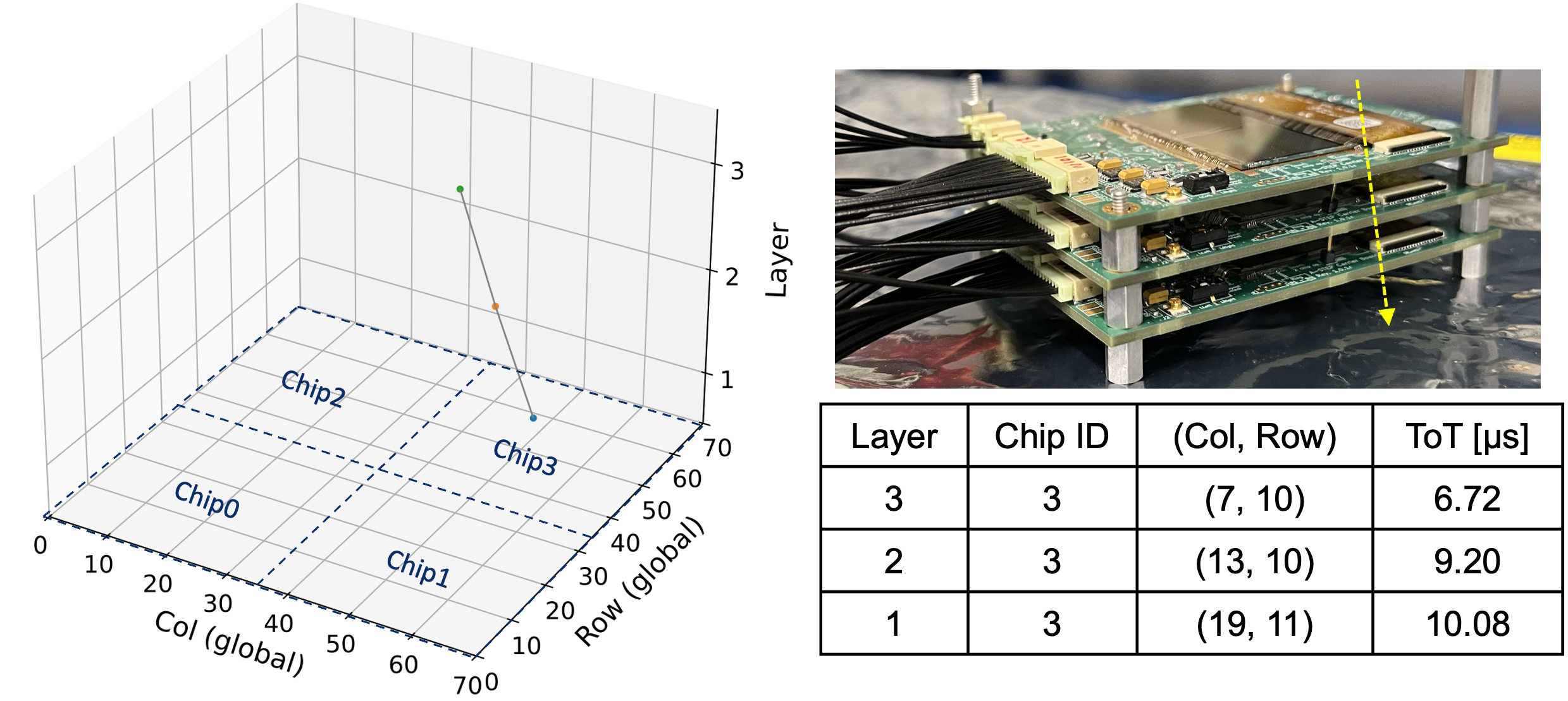}
    \caption{Cosmic-ray event display of a three-layer stack of quad-chips (left) and hit information.}
    \label{fig:3q}
\end{figure}

The cosmic-ray test was performed to verify the performance of a three-layer stack of uncalibrated quad-chips and to ensure the proper operation of the software and firmware under development. 
%This system was designed to share the AstroPix timestamp clock across the three quad-chip layers. 
The system was designed to distribute a common AstroPix ToA clock across the three quad-chip layers to enable coincident measurements.

Fig.~\ref{fig:3q} shows a representative event display of a cosmic-ray trajectory reconstructed from a coincident event in which all three layers recorded the same AstroPix ToA within one tick, corresponding to 400~ns. 
In the figure, the top-right panel shows the three-layer quad-chip configuration, and the bottom-right table presents the corresponding hit information.
Since the chips have not yet been calibrated, the ToT values were not considered in the analysis.
%The three plots on the right show the ToT values at the hit positions in each quad-chip for this coincident event. 
%These ToT maps illustrate the measured cosmic-ray energy deposition in the pixels, while also providing information about the hit positions.

The chips in each quad-chip module are daisy-chained such that data from chip 3 is sequentially transmitted to chip 0 and then forwarded to the FPGA. As the three quad-chip layers operate simultaneously, this demonstrates successful readout of 12 chips in total. These results confirm reliable system communication and validate the capability of the prototype A-STEP detector to operate multiple daisy-chain configurations with shared timing information.
%As the daisy-chained readout operates sequentially from chip~0 to chip~3, this demonstrates that the readout of all three quad-chip layers corresponds to a total of 12 single-chips being successfully read out. This result confirms reliable system communication and validates the capability of the prototype A-STEP detector to read out data from all quad-chips, up to 12 chips in total.

%This plot shows a coincident event in chip~3 across all layers. 
%The daisy-chained readout operates sequentially from chip~0 to chip~3, demonstrating that the readout of all chip~3s corresponds to a total of 12 chips being read out. 
%This result confirms that the system establishes reliable communication and validates the capability of the prototype A-STEP detector to read out data from all quad-chips, up to 12 chips in total.
%This result demonstrates that the system establishes reliable communication and validates the capability of the prototype A-STEP detector to read out data from all quad-chips, up to 12 chips in total.

\subsection{Nine-chip prototype module as a unit of the BIC imaging layer}
%Initial functionality confirmed – power stable and DAQ communication OK. 
%\begin{figure}[h!]
%    \centering
%    \includegraphics[width=0.9\linewidth]{figs/9_noise.png}
%    \caption{Mask map of 9-chip module as a result of noise scan}
%    \label{fig:9n}
%\end{figure}
%Fig.~\ref{fig:9n} shows the mask maps of a nine-chip module.
%In these maps, masked pixels are displayed in black, 

\begin{figure}[h!]
    \centering
    \includegraphics[width=1\linewidth]{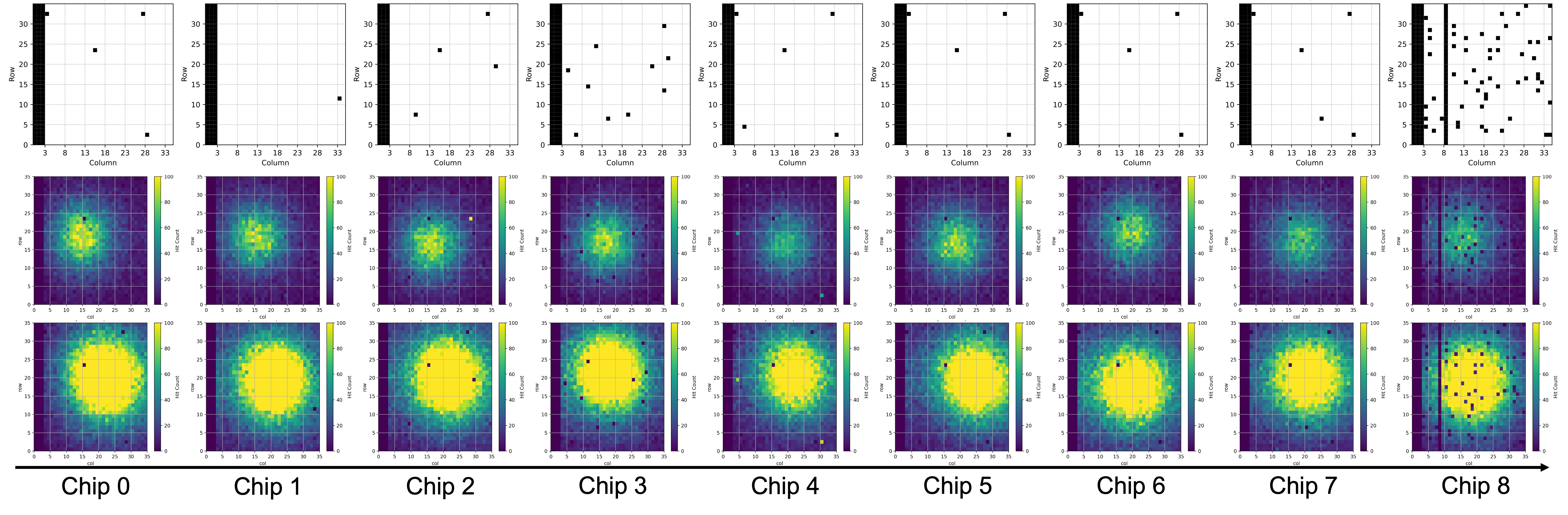}
    \caption{Mask map of the nine-chip module (top) and hit maps from the $^{90}$Sr source test with the nine-chip module using 3~mm (middle) and 5~mm collimators (bottom).}
    \label{fig:9srhit}
\end{figure}

The noise scan for the nine-chip prototype module was performed in the same way as for the quad-chip test, and the resulting mask maps for each chip are shown in Fig.~\ref{fig:9srhit}, similar to those in Fig.~\ref{fig:qnoise}.
In the nine-chip module, a tighter noise-rate cut of 1 Hz was applied, compared to 2 Hz for the quad-chip module.
The current prototype achieved a 99\% active pixel yield for all chips except the last chip, which had 91 noisy pixels corresponding to a 91.9\% active pixel yield. 
In pre-production, only chips that pass quality control with more than 99\% good pixels will be used.
\begin{figure}[h!]
    \centering
    \includegraphics[width=0.47\linewidth]{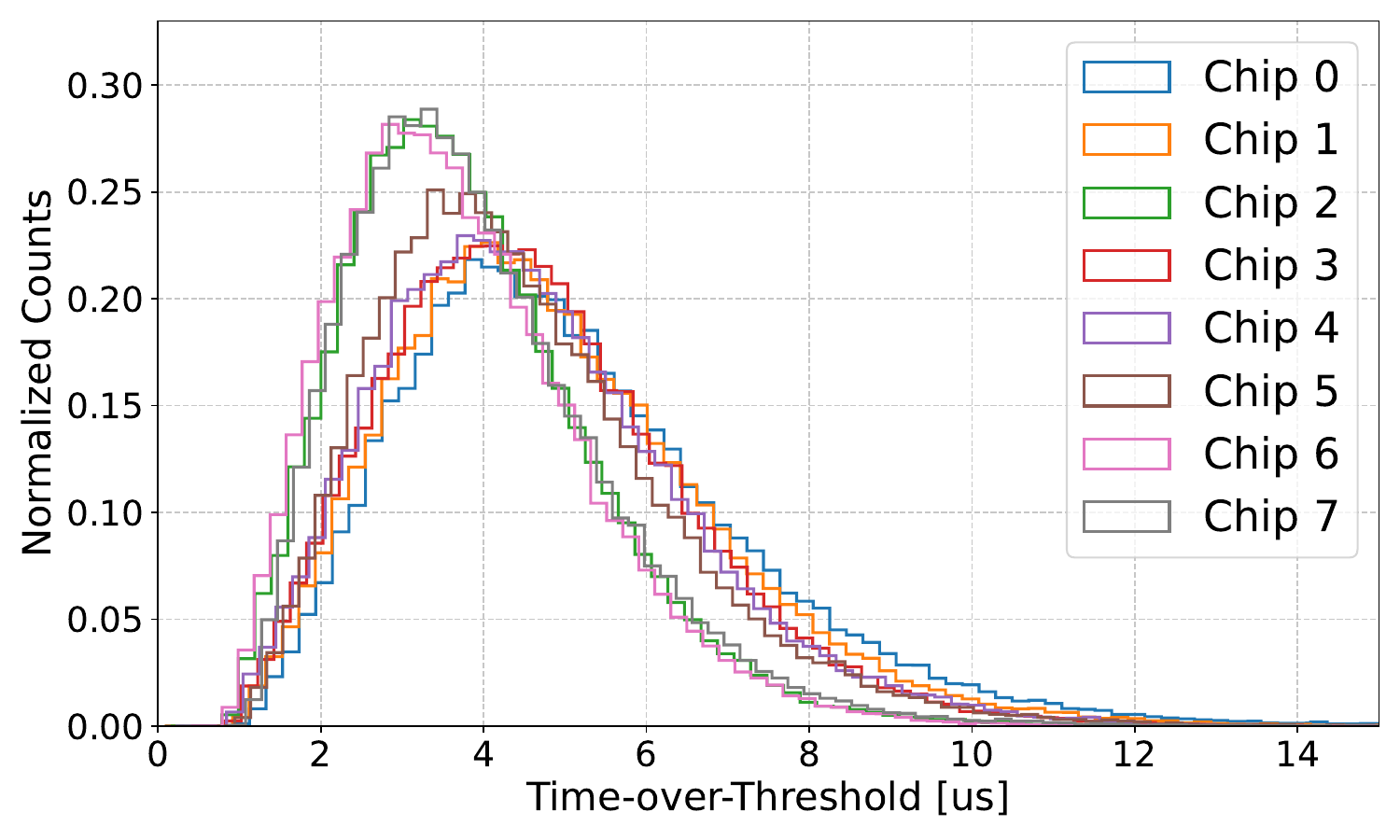}
    \caption{ToT distributions from the $^{90}$Sr source test with the nine-chip module}
    \label{fig:9srtot}
\end{figure}

Using this module, a source test was subsequently conducted. 
A 10~\unit{\micro Ci} $^{90}$Sr source was manually moved to the next chip every 1 minute. 
Fig.~\ref{fig:9srhit} shows hit maps for each chip obtained with 3 and 5 mm diameter collimators, where black pixels indicate masked pixels. 
The hit maps exhibit good alignment with the source position, with a larger number of responsive pixels and higher hit counts observed for the larger-diameter collimator.

The normalized ToT distributions for chips 0--7, excluding chip 8, are presented in Fig.~\ref{fig:9srtot}, obtained from the $^{90}$Sr source measurement using a 5 mm diameter collimator.
They are well described by a Landau function convoluted with a Gaussian function.  
Each distribution includes all pixels within the corresponding chip and reflects its average response characteristics.
The nine-chip module exhibits two distinct ToT response groups: one comprising chip 2, 6, and 7, which show a relatively lower ToT response with a most probable value around 3$~\unit{\micro s}$, and another including chip 0, 1, 3, 4, and 5, which exhibit a slightly higher ToT response with a most probable value around 4$~\unit{\micro s}$.
In addition, consistent ToT distributions were observed regardless of the collimator size, and the distributions are comparable to those measured in the single-chip tests.
These results demonstrate that the nine-chip prototype module operated as intended with the current setup, including daisy-chained readout.

\begin{figure}[h!]
    \centering
    \includegraphics[width=0.47\linewidth]{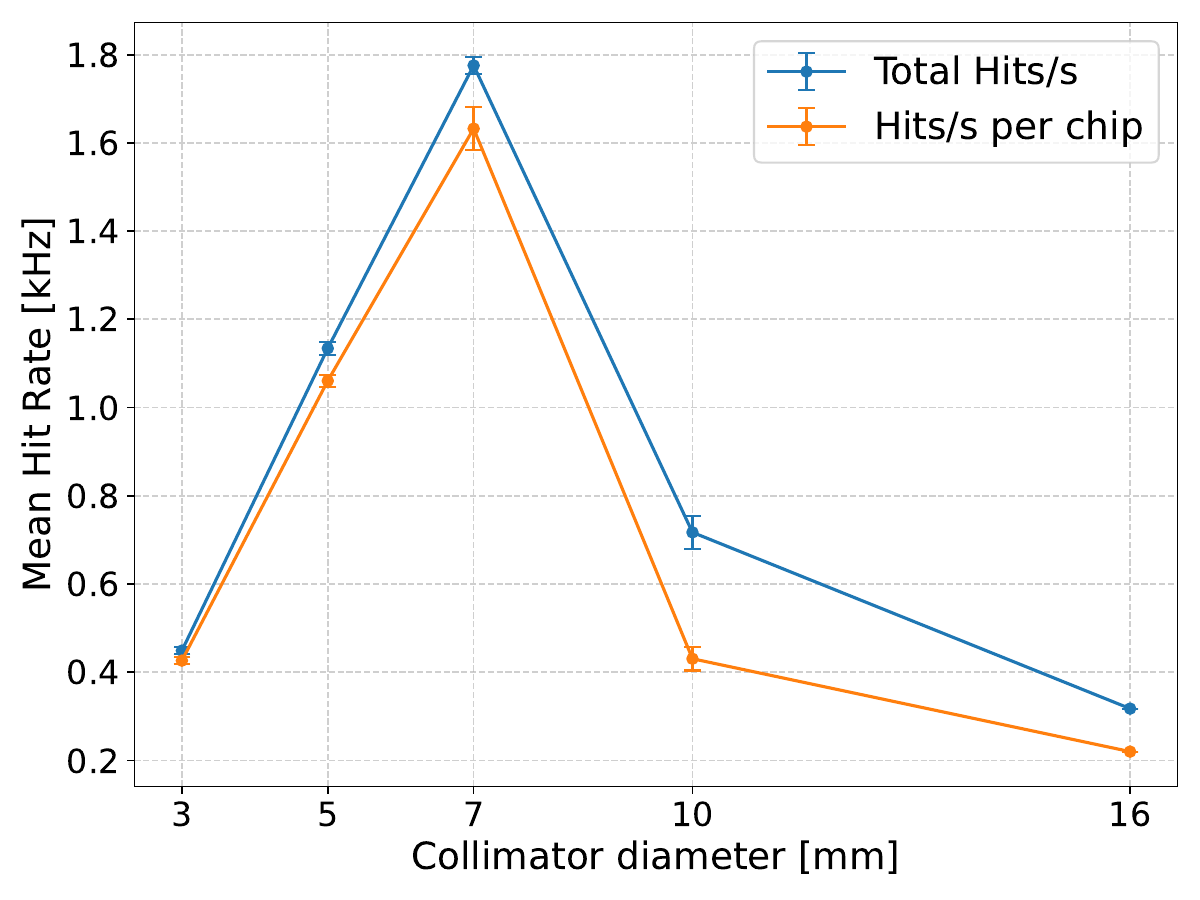}
    \caption{Mean hit rates per nine-chip module and per chip measured in the $^{90}$Sr source test.}
    \label{fig:9rate}
\end{figure}
With the same setup, a hit rate measurement was performed to validate the readout capability of the nine-chip module and to measure the maximum achievable hit rate. 
A 10~\unit{\micro Ci} $^{90}$Sr source with different collimators was placed above chip~1, which had only one masked pixel.
The hit rate, defined as the number of recorded hits divided by the data-taking time, was measured five times for each point, and the mean values and standard deviations were plotted in Figure~\ref{fig:9rate}.
The figure shows the mean hit rates as a function of collimator size, with the total and per-chip rates indicated in blue and orange, respectively.
The maximum per-chip rate was measured to be about 1.6 kHz, and reliable operation was maintained up to a collimator diameter of 7 mm, beyond which significant data dropout were observed with the current setup.

The maximum expected hit rate per chip for the imaging layers in the EIC environment is about 925 Hz, taking into account DIS, electron, and proton beam backgrounds.
For space-mission conditions, the expected hit rate of the A-STEP detector is approximately 1~\unit{Hz/cm^{2}}, and under the assumption of a very bright GRB simulation, it increases to about 45 Hz per chip for AMEGO-X.
Under the current setup, the nine-chip prototype module satisfies the requirements for both the BIC imaging layers and A-STEP.
%Move to summary below sentences
%These measurements indicate that operation under realistic beam conditions is promising.
%Although the hit-rate measurement was performed with the nine-chip prototype, the results demonstrate that the performance is sufficient to meet the requirements of the quad-chip detector intended for the A-STEP space mission as well.
%The maximum measured hit rate per pixel is about 3.6 Hz, which reflects the FIFO limitation of the AstroPix\_v3 design, while starting with the AstroPix4 design, an individual FIFO is implemented for each pixel.
%We expect that hit rate per pixel will improved in AstroPix4.

\section{Summary}
%Several AstroPix\_v3 configurations, including the quad-chip and the nine-chip prototype modules, were successfully tested.
%The quad-chip achieved a high active-pixel yield of about 99\%, with stable powering and reliable daisy-chained readout.
%A three-layer stack of quad-chips demonstrated synchronized multi-layer operation, validating the scalability of the design.
%The nine-chip prototype further confirmed stable operation and achieved hit-rate measurements consistent with expectations.
%The measured performance satisfies the requirements for both the BIC imaging layers at the EIC and the A-STEP space mission.
%These measurements indicate that operation under realistic beam conditions is promising.
%Although the hit-rate test was performed with the nine-chip prototype, the results demonstrate that the performance is sufficient to meet the requirements of the quad-chip detector intended for the A-STEP mission as well.
%Overall, the AstroPix\_v3 sensor design has proven feasible for applications in both collider and space-based environments.
Several AstroPix\_v3 configurations, including the quad-chip and nine-chip prototype modules, were successfully tested.
The quad-chip achieved reliable readout and a three-layer stack demonstrated synchronized multi-layer operation, confirming the scalability of the design.
The nine-chip prototype also achieved a 99\% active-pixel yield with stable performance and satisfied the hit-rate requirements, indicating reliable operation under realistic beam conditions.
The measured performance meets the requirements for both the BIC imaging layers at the EIC and the A-STEP space mission.
Although the hit-rate test was conducted with the nine-chip prototype, the results confirm sufficient performance for the quad-chip detector planned for A-STEP.
Overall, the AstroPix\_v3 design proves feasible for both collider and space-based detector applications, while the current design is limited to to a maximum hit rate of approximately 4~Hz per pixel.
Future versions are under development to achieve substantially higher hit rates at both the pixel and chip levels, along with full depletion, fast time resolution, and low power consumption.

\end{document}